\documentclass[a4paper,11pt]{article}
\usepackage[T1]{fontenc}
\usepackage[latin1]{inputenc}
\usepackage[english]{babel}
\usepackage{amsmath,amsfonts,amssymb,amsthm}
\usepackage{graphicx}
\usepackage{fancyvrb}
\usepackage{url}
\usepackage{nicefrac}
\usepackage{vmargin}
\usepackage{feynmp}
\usepackage{cancel}
\usepackage{verbatim}
\usepackage{fancyhdr}

\setmarginsrb{90pt}{25pt}{95pt}{45pt}{35pt}{23pt}{12pt}{35pt}

\begin{document}
\title{$\phantom{A}$\\New Relations for Gauge-Theory Amplitudes with Matter}


\author{Thomas S\o{}ndergaard\footnote{email: tsonderg@nbi.dk} \\ \\
\small \textit{The Niels Bohr International Academy, The Niels Bohr Institute,} \\ \small \textit{Blegdamsvej 17, DK-2100 Copenhagen, Denmark} }


\date{}
\maketitle

\begin{abstract}
We extend a recently discovered set of relations for gauge-theory amplitudes to non-gluonic matter. For all MHV amplitudes we find that these can be made to hold for scalar/fermion/quark cases by inclusion of a factor derived via Ward identities. For six- and seven-point amplitudes with non-gluonic matter we explicitly confirm these relations for NMHV helicity configurations.
\end{abstract}

\thispagestyle{fancy}
\section{\label{intro}Introduction}
In recent years new efficient methods for scattering amplitude calculations have emerged. Witten's proposed duality between tree-level amplitudes and twistor
string theory \cite{Witten:2003nn}, became the inspiration for a novel technique, today known as the CSW formalism \cite{Cachazo:2004kj}. This approach uses MHV \cite{Parke:1986gb,Berends:1987me} amplitudes as building blocks for
more complicated amplitudes. Shortly after a related on-shell recursion
relation was introduced (BCFW) \cite{Britto:2004ap,Britto:2005fq}. 

Beside the mere practical calculational advantages, these techniques
have lead to interesting scientific discoveries. The BCFW recursion
relations have been used to derive a set of consistency conditions for
the S-matrix for massless particles, providing non-trivial constraints
on scattering amplitudes \cite{Benincasa:2007xk,Schuster:2008nh}. 

A number of useful relations between color-ordered tree amplitudes have also been known for some time. These include those directly
dictated by the color-group \cite{Mangano:1990by,Dixon:1996wi} as well
as the Kleiss-Kuijf relations \cite{Kleiss:1988ne,DelDuca:1999rs}. The
latter reduce the number of independent partial $n$-point gluon
amplitudes to $(n-2)!$ In addition supersymmetry provide relations between amplitudes of equal helicity configuration but different particle content, known as supersymmetric Ward identities (SWI) \cite{Grisaru:1976vm,Grisaru:1977px,Parke:1985pn}. 

Recently new relations for color-ordered tree amplitudes have been conjectured. Based on a kinematic identity Bern, Carrasco and
Johansson \cite{Bern:2008qj} have presented a remarkable set of helicity-independent relations for gluonic amplitudes, reducing the number of
independent partial amplitudes even further to $(n-3)!$

In this paper we make a non-trivial extension of these new gluonic
relations to color-ordered tree amplitudes containing matter. We will consider amplitudes with $n-2$ gluons and two adjoint fermions (gluinos), adjoint scalars or a quark-antiquark pair. The extended relations introduce a sign-function due to fermi statistics, and are verified for $n$-point MHV amplitudes and explicitly checked in NMHV cases for six and seven point amplitudes. 


\section{\label{first.sec} Gauge-theory relations}
In this section we introduce supersymmetric Ward identities and matter MHV amplitudes \cite{Nair:1988bq} in terms of gluon MHV amplitudes. We also briefly review the newly discovered gluonic relations. 

Our conventions and notation can be found in appendix A.

\subsection{\label{SWI}Supersymmetry relations and MHV amplitudes}
As is well known in a supersymmetric theory, particles with different helicity are related to each other by a symmetry transformation. These transformations can be used to obtain relations between amplitudes with different particle content, known as supersymmetric Ward identities.

The identities arise from the fact that the supercharge, $Q$, annihilates the vacuum. This imply that
\begin{eqnarray}
0 = \langle 0|\lbrack Q, \Phi_1\Phi_2\cdots\Phi_n\rbrack|0\rangle = \sum_{i=1}^n\langle 0|\Phi_1\cdots\lbrack Q,\Phi_i\rbrack\cdots\Phi_n|0\rangle,
\label{general_SWI}
\end{eqnarray}
where $\Phi_i$ create helicity eigenstates of some specific particle type. To get concrete relations one need to know the commutators between the supercharge $Q$ and the $\Phi_i$'s.  

As an example consider the $\mathcal{N}=1$ supersymmetric Yang-Mills theory. The operators that create gluons and gluinos of helicity $\pm$ with momentum $p$ are denoted by $g^{\pm}(p)$ and $\Lambda^{\pm}(p)$, respectively. They obey the following commutator relations
\begin{eqnarray}
&&\!\!\!\!\!\lbrack Q(\eta),g^{+}(p)\rbrack = -\Gamma^{+}(p,\eta)\bar{\Lambda}^{+}(p), \qquad
\lbrack Q(\eta),g^{-}(p)\rbrack = \Gamma^{-}(p,\eta)\Lambda^{-}(p), \nonumber \\
&&\!\!\!\!\!\lbrack Q(\eta),\bar{\Lambda}^{+}(p)\rbrack = -\Gamma^{-}(p,\eta)g^{+}(p), \qquad
\lbrack Q(\eta),\Lambda^{-}(p)\rbrack = \Gamma^{+}(p,\eta)g^{-}(p),
\end{eqnarray}
where $Q$ depends on a fermionic spinor parameter $\eta$. We can always choose a $\eta$ so that $\Gamma^{\pm}$ is given by 
\begin{eqnarray}
\Gamma^+(p,q) = \theta\lbrack qp\rbrack, \qquad \Gamma^-(p,q) = \theta\langle qp\rangle,
\end{eqnarray}
with $q$ being an arbitrary massless vector and $\theta$ a Grassmann variable (which will drop out of the calculations) \cite{Mangano:1990by,Dixon:1996wi}.

It turns out that the identities with MHV amplitudes are especially simple. Take for instance the string of operators $g_1^-\bar{\Lambda}_2^+g_3^-g_4^+\cdots g_n^+$ in Eq.~(\ref{general_SWI}), leading to the following SWI
\begin{eqnarray}
0 \!\!\!&=&\!\!\! \Gamma^-(1,q)A_n(\Lambda_1^-,\bar{\Lambda}_2^+,g_3^-,g_4^+,\ldots,g_n^+) - \Gamma^-(2,q)A_n(g_1^-,g_2^+,g_3^-,g_4^+,\ldots,g_n^+) \nonumber \\
&&\!\!\!-\Gamma^-(3,q)A_n(g_1^-,\bar{\Lambda}_2^+,\Lambda_3^-,g_4^+,\ldots,g_n^+),
\end{eqnarray}
where we omit amplitudes with two gluinos of the same helicity since these vanish due to helicity-conservation. Note that we have an additional sign change in front of the last amplitude due to the Grassmannian nature of $\Gamma^{\pm}$ when moving it through $\Lambda_2$. Choosing $q=p_3$ the last contribution vanishes, and we obtain
\begin{eqnarray}
A_n(\Lambda_1^-,\bar{\Lambda}_2^+,g_3^-,g_4^+,\ldots,g_n^+) = \frac{\langle 32\rangle}{\langle 31\rangle} A_n(g_1^-,g_2^+,g_3^-,g_4^+,\ldots,g_n^+).
\label{MHV_example}
\end{eqnarray}

The N$^k$MHV ($k\geq 1$) relations are much more complex and do not provide similarly simple relations like Eq.~(\ref{MHV_example}).

Since we focus on amplitudes with adjoint fermions or adjoint scalars, an $\mathcal{N}\geq 2$ supersymmetric theory would be needed to derive the relevant relations. Examples of various SWI calculations can be found in \cite{Mangano:1990by,Dixon:1996wi,Parke:1985pn,Mangano:1987kp,Bidder:2005in,Britto:2005ha}.
We will denote the particle type by a subscript on the corresponding momentum index. The subscript $s$ denotes adjoint scalars, $f$ (and $\bar{f}$) adjoint fermions, and no subscript denotes gluons. We will use the following MHV matter amplitudes, which can be obtained by SWI calculations similar to Eq.~(\ref{MHV_example}), (for further details see for instance \cite{Georgiou:2004wu,Glover:2008tu})
\begin{eqnarray}
m_n(i^-,j^-_{f/s},k^+_{\bar{f}/s}) = \left(\pm \frac{\langle ik \rangle}{\langle ij \rangle}\right)^{2-2h} A_n(i^-,j^-), \qquad
A_n(i^-,j^-) = \frac{\langle ij \rangle^4}{\prod_{r=1}^n\langle r\,(r+1)\rangle}.
\label{MHV} 
\end{eqnarray}
Here we omit writing the positive-helicity gluons, and inside the parenthesis we have a $+$ when $j<k$ and $-$ when $k<j$. $A_n$ is the purely gluonic amplitude, and the (SWI) factor has $h=0$ for scalars and $h=\nicefrac{1}{2}$ for fermions (and $h=1$ for gluons). Note that the $+$ ($-$) on a scalar is to be understood as particle (antiparticle).

Finally we will make use of an identity \cite{Mangano:1987kp} between color-ordered tree amplitudes with fermions in the fundamental representation (quarks) and color-ordered tree amplitudes with fermions in the adjoint representation (gluinos). We denote a quark by the subscript $q$ and an antiquark by $\bar{q}$ 
\begin{eqnarray}
m_n(1_{\bar{q}},2_q,3,\ldots,n) = m_n(1_{\bar{f}},2_f,3,\ldots,n),
\label{quark_fermion}
\end{eqnarray}
where $1_{\bar{q}}=1_{\bar{f}}$ and $2_q=2_f$.

\subsection{New gluonic relations}
In \cite{Bern:2008qj} a new representation of gluonic tree amplitudes are conjectured. This representation introduces an identity satisfied by kinematic numerator factors, analogous to the Jacobi identity for color-factors. The kinematic identity constrain the partial amplitudes and lead to new non-trivial relations.

The conjecture can be realized as an equation system describing the unknown numerators $n_i$
\begin{eqnarray}
 \{ n_{\alpha} = n_{\beta} - n_{\gamma} \}, \qquad
A_n(\sigma_i\{1,2,3,\ldots,n\}) = \left[ \sum_j\frac{n_j}{(\prod_mp^2_m)_j} \right]_i,
\label{conj_1}
\end{eqnarray}
where the first equation represents all the numerator identities analogous to the Jacobi identities for the corresponding color-factors, and the second is the statement that the terms in this representation sum up to the known partial amplitudes for at least $i= 1,\ldots,(n-3)!$ different permutations of the external legs (the \textit{basis} amplitudes). The product of $p^2_m$'s in the denominator describe the pole-structure of the amplitude. 

It should be noted, that there is a certain amount of freedom in the solution to the above equations. However, part of the conjecture is that any amplitude built out of the solution to Eq.~(\ref{conj_1}) will be independent of this freedom.

We will illustrate the above conjecture in the simplest case (see also the earlier work about certain zeros in cross sections \cite{Zhu:1980sz}). At four points the new representation imply that the full color-dressed tree-level gluon amplitude can be written as
\begin{eqnarray}
\mathcal{A}_4^{\mathrm{tree}} = g^2\left( \frac{c_sn_s}{s} + \frac{c_tn_t}{t} + \frac{c_un_u}{u} \right),
\end{eqnarray}
where $s$, $t$ and $u$ are the usual Mandelstam variables, $c_i$ are
color-factors satisfying the Jacobi identity $c_u = c_s - c_t$, and the
$n_i$ are kinematic factors satisfying the corresponding identity $n_u =
n_s - n_t$ (Eq.~(\ref{conj_1})).

We will choose our basis amplitude to be $A_4(1,2,3,4)$, and following the second statement in Eq.~(\ref{conj_1}) we get
\begin{eqnarray}
A_4(1,2,3,4) = \frac{n_s}{s} + \frac{n_t}{t}.
\end{eqnarray}
 
In order to obtain the mentioned relation, we write $A_4(1,2,4,3)$ in this representation as well
\begin{eqnarray}
A_4(1,2,4,3) = -\frac{n_u}{u} - \frac{n_s}{s},
\end{eqnarray}
where the sign flip is due to the antisymmetry of color-ordered Feynman rules.

By use of the kinematic identity $n_u = n_s - n_t$ we get
\begin{eqnarray}
tA_4(1,2,3,4)-uA(1,2,4,3)\!\!\! &=&\!\!\! \frac{tn_s}{s} + (n_t + n_u) + \frac{un_s}{s}  \nonumber \\
&=&\!\!\! \frac{tn_s}{s} + n_s + \frac{un_s}{s} = \frac{n_s(t+s+u)}{s} = 0,
\end{eqnarray}
hence we see that we derived the well known identity $tA_4(1,2,3,4)=uA(1,2,4,3)$. 

In general, choosing the basis as $A_n(1,2,3,\sigma\{4,\ldots,n\})$ and solving the above equation system for these $(n-3)!$ gluon amplitudes, \textit{i.e.} expressing $n_i$ in terms of these amplitudes, and using the solution for the remaining amplitudes, one obtains new relations equivalently to the four-point case. 

It is possible to conjecture an all-$n$ form \cite{Bern:2008qj} for these gluonic relations,
\begin{eqnarray}
A_n(1,2,\{\alpha\},3,\{\beta\})=
\sum_{\{\sigma\}\in POP(\{\alpha\},\{\beta\})} A_n(1,2,3,\{\sigma\})\prod_{k=4}^m\frac{\mathcal{F}(3,\{\sigma\},1|k)}{s_{2,4,\ldots,k}},\phantom{a}
\label{BCJ}
\end{eqnarray}
where
\begin{eqnarray}
\{\alpha\}  \equiv  \{4,5,\ldots,m-1,m\},\qquad
\{\beta\}  \equiv  \{m+1,m+2,\ldots,n-1,n\},
\end{eqnarray}
and the sum runs over ``partially ordered permutations'' ($POP$), corresponding to all permutations of $\{\alpha\}\cup\{\beta\}$ that maintains the order of the $\{\beta\}$ elements. Either set can be taken as empty, but if $\{\alpha\}$ is empty the relation becomes trivial. The function $\mathcal{F}$ associated with leg $k$ is given by
\begin{eqnarray}
\mathcal{F}(3,\{\sigma\},1|k) \equiv \mathcal{F}(\{\rho\}|k)  =  
\left\{ \begin{array}{ll}
\sum_{l=t_k}^{n-1}\mathcal{G}(k,\rho_l) & \textrm{if $t_{k-1}<t_k$}  \\
-\sum_{l=1}^{t_k}\mathcal{G}(k,\rho_l) & \textrm{if $t_{k-1}>t_k$}       \end{array} \right\}\phantom{AAAAAAAAAA} \nonumber \\
  + \left\{ \begin{array}{ll} 
s_{2,4,\ldots,k} & \textrm{if $t_{k-1}<t_k<t_{k+1}$} \\
-s_{2,4,\ldots,k} & \textrm{if $t_{k-1}>t_k>t_{k+1}$} \\
0 & \textrm{else}         \end{array}  \right\},
\end{eqnarray}
where $t_k$ is the position of leg $k$ in $\{\rho\}$, except for $t_3$ and $t_{m+1}$ which are always defined to be
\begin{eqnarray}
t_3 \equiv t_5, \qquad t_{m+1} \equiv 0,
\end{eqnarray}
and $\rho_l$ is the leg at position $l$ in $\{\rho\}$. The function $\mathcal{G}$ is given by
\begin{eqnarray}
\mathcal{G}(i,j) = \left\{ \begin{array}{ll} 
s_{i,j} & \textrm{if $i<j$ or $j=1,3$} \\
0 & \textrm{else}
                           \end{array}   \right\},
\end{eqnarray}
and the kinematic invariants are,
\begin{eqnarray}
s_{i,j} = (k_i+k_j)^2, \qquad s_{2,4,\ldots,i} = (k_2+k_4+\ldots+k_i)^2,
\end{eqnarray}
with all momenta massless and outgoing.

The ordering in $\{\alpha\}$ and $\{\beta\}$ only give some of the
relations, but the rest can be obtained by permutations of $4,\ldots,n$. 

Note that the Kleiss-Kuijf relations allow for the fixing of leg 1
and 2, and that the basis amplitudes have been chosen to be
independent under these relations. Therefore only $(n-3)!$ independent partial amplitudes are left. 

For purely gluonic amplitudes the four- and five-point relations, generated from Eq.~(\ref{BCJ}), are \cite{Bern:2008qj}
\begin{eqnarray}
A_4(1,2,\{4\},3) = \frac{A_4(1,2,3,4)s_{14}}{s_{24}},
\label{4_BCJ}
\end{eqnarray}
and
\begin{eqnarray}
&&\!\!\!A_5(1,2,\{4\},3,\{5\}) = \frac{A_5(1,2,3,4,5)(s_{14}+s_{45}) + A_5(1,2,3,5,4)s_{14}}{s_{24}},\phantom{A} \nonumber \\
&&\!\!\!A_5(1,2,\{4,5\},3) = \frac{-A_5(1,2,3,4,5)s_{34}s_{15}
-A_5(1,2,3,5,4)s_{14}(s_{245}+s_{35})}{s_{24}s_{245}},
\label{5_BCJ}
\end{eqnarray}
respectively. 

In the four-point case only one relation exists. In the five-point case two more relations are obtained by interchanging 4 and 5 in the two equations. We have kept $\{\dots\}$ in the equations for easy comparison with Eq.~(\ref{BCJ}). Explicit expressions for six- and seven-point relations can be found in appendix B.

In the following we will refer to the above new relations as the \textit{BCJ-relations}.

\section{New matter relations}
Our conjecture is that Eq.~(\ref{BCJ}) can be generalized to include matter, and thereby lead to new non-trivial relations between matter amplitudes. 

Let $i$ and $j$ denote the momentum index of the two helicity $h$ matter particles in the partial amplitudes $m_n$. Our generalized formula is  
\begin{eqnarray}
m_n(1,2,\{\alpha\},3,\{\beta\})=
\sum_{\{\sigma\}\in POP(\{\alpha\},\{\beta\})} \big(\mathrm{sign}_{\sigma}\!\{ i,j\}\big)^{2h}\, m_n(1,2,3,\{\sigma\})\prod_{k=4}^m\frac{\mathcal{F}(3,\{\sigma\},1|k)}{s_{2,4,\ldots,k}},
\label{BCJ_matter}
\end{eqnarray}
where $\mathrm{sign}_{\sigma}\!\{ i,j\}$ is a sign-function defined to give $-1$ when the order of the matter legs in $m_n(1,2,3,\{\sigma\})$ is changed compared to $m_n(1,2,\{\alpha\},3,\{\beta\})$, and $+1$ when the order is kept. We have $h=0$ for scalars, $h=\nicefrac{1}{2}$ for fermions (gluinos or quarks), and $h=1$ for gluons. Note that the quark case assume $(i,j)=(1,2)$. 

We now go into details concerning the verification of these extended relations, which are based on the simple structure of the MHV matter amplitudes from Sec.~\ref{SWI} and explicit checks in six- and seven-point NMHV cases. We stress that all the amplitudes used in the verification are consistent with the supersymmetric Ward identities.

\subsection{Four-point matter relation}
For four-point amplitudes only MHV configurations are non-zero. It is evident from Eq.~(\ref{MHV}) that Eq.~(\ref{4_BCJ}) can be extended to all MHV matter amplitudes with two fermions or two scalars. One simply multiplies both side of Eq.~(\ref{4_BCJ}) with the appropriate SWI factor and get, for instance
\begin{eqnarray}
m_4(1^{-}_{s/f},2^{+}_{s/\bar{f}},\{4^-\},3^+) =
\frac{m_4(1^{-}_{s/f},2^{+}_{s/\bar{f}},3^+,4^-)s_{14}}{s_{24}}.
\label{4_ex}
\end{eqnarray}
If we consider the case with fermions on leg 3 and 4 we see that one of the sides need to change sign in order for the relation to hold, \textit{e.g.} after multiplying with the same SWI factor on both sides of Eq.~(\ref{4_BCJ}) we rewrite the righthand side amplitude as
\begin{eqnarray}
\frac{\langle 13\rangle}{\langle 14\rangle}A_4(1^-,2^+,3^+,4^-) = -\left(-\frac{\langle 13\rangle}{\langle 14\rangle}A_4(1^-,2^+,3^+,4^-) \right) = - m_4(1^-,2^+,3_{\bar{f}}^+,4_f^-).
\end{eqnarray}

Furthermore Eq.~(\ref{quark_fermion}) imply that also quark amplitudes $m_4(1_{\bar{q}},2_q,3,4)$ satisfy the relation.

\subsection{Five-point matter relations}
Again only MHV (and googly-MHV) amplitudes are non-zero, and
Eq.~(\ref{MHV}) allow
us to write down the analogous matter relations. The example similar to Eq.~(\ref{4_ex})
is
\begin{eqnarray}
m_5(1^{-}_{s/f},2^{+}_{s/\bar{f}},\{4,\},3,\{5\})\!\!\! &=&\!\!\! \frac{m_5(1^{-}_{s/f},2^{+}_{s/\bar{f}},3,4,5)(s_{14}+s_{45})}{s_{24}}
+ \frac{m_5(1^{-}_{s/f},2^{+}_{s/\bar{f}},3,5,4)s_{14}}{s_{24}},\nonumber \\
m_5(1^{-}_{s/f},2^{+}_{s/\bar{f}},\{4,5\},3)\!\!\! &=&\!\!\! -\frac{m_5(1^{-}_{s/f},2^{+}_{s/\bar{f}},3,4,5)s_{34}s_{15}}{s_{24}s_{245}}  \nonumber \\
&&\phantom{AAAAAAAAA}-\frac{m_5(1^{-}_{s/f},2^{+}_{s/\bar{f}},3,5,4)s_{14}(s_{245}+s_{35})}{s_{24}s_{245}}.
\end{eqnarray}
If the fermion legs were among those that flip order in the amplitudes (\textit{i.e.} if they both belong to $\{3,4,5\}$) the same argument as in the four-point case verify the sign function in Eq.~(\ref{BCJ_matter}). 

Eq.~(\ref{quark_fermion}) implies the validity of these relations for quark amplitudes $m_5(1_{\bar{q}},2_q,3,4,5)$ as well.

\subsection{$n$-point MHV matter case}
The four- and five-point cases are just specific examples of the general MHV $n$-point situation. The simple multiplication by a SWI factor and the linearity of the BCJ-relations extend Eq.~(\ref{BCJ}) directly to include scalars or fermions in the manner conjectured in Eq.~(\ref{BCJ_matter}), \textit{e.g}. 
\begin{eqnarray}
m_n(1^{-}_{s/f},2,\{\alpha\},3^{+}_{s/\bar{f}},\{\beta\})= 
\sum_{\{\sigma\}\in POP(\{\alpha\},\{\beta\})} m_n(1^{-}_{s/f},2,3^{+}_{s/\bar{f}},\{\sigma\}) \prod_{k=4}^m\frac{\mathcal{F}(3,\{\sigma\},1|k)}{s_{2,4,\ldots,k}}.
\end{eqnarray}
When both fermions belong to $\{3,\ldots,n\}$ the relations involve amplitudes flipping the order of these legs, which is accounted for by the sign-function. 

The relations are extended to MHV amplitudes with fermions in the fundamental representation by Eq.~(\ref{quark_fermion}).

\subsection{Six- and seven-point matter relations}
For six-point amplitudes there are three classes of matter relations, giving a total of 18 relations obtained by permutations of leg 4, 5 and 6. 

For seven-point amplitudes there are 96 matter relations. They come in four classes, each representing 24 relations by permutating 4, 5, 6 and 7. We have generated these four classes from Eq.~(\ref{BCJ}). Explicit expression for both six- and seven-point relations can be found in appendix B.

Until now the verification of Eq.~(\ref{BCJ_matter}) has been provided
by the multiplication of an SWI factor. The NMHV amplitudes, however,
introduce more complicated expressions. We have used the
BCFW recursion relation to generate NMHV matter amplitudes for six and seven points. When possible, we have compared with amplitudes obtained in the literature \cite{Britto:2004ap,Britto:2005ha,de_Florian:2006ek}, and made sure that all used amplitudes were consistent with supersymmetric Ward identities. At tree-level the following amplitude structure appears
\begin{eqnarray}
A_n^{\mathrm{gluon}} =  \sum_i X_i, \qquad m_n = \sum_i X_i\,(a_i)^{2-2h},
\end{eqnarray}
with $m_n$ containing a pair of helicity $h$ matter particles, and $a_i$ is some kinematical (SWI) factor. An example of a matter amplitude is
\begin{eqnarray}
m_6(1_{\bar{h}}^+,2^+,3^+,4_h^-,5^-,6^-) = \frac{\langle 6|1+2|3\rbrack^3}{\langle 12\rangle \langle 61\rangle \lbrack 34\rbrack \lbrack 45\rbrack \langle 2|1+6|5\rbrack s_{126}} \left( \frac{\lbrack 34 \rbrack \langle 16\rangle}{ \langle 6|1+2|3\rbrack} \right)^{2-2h}  \nonumber \\
 \frac{\langle 4|2+3|1\rbrack^3}{\langle 23\rangle \langle 34\rangle \lbrack 56\rbrack \lbrack 61\rbrack \langle 2|3+4|5\rbrack s_{234}} \left( \frac{s_{234}}{\langle 4|2+3|1\rbrack} \right)^{2-2h}.
\label{6pt_amp}
\end{eqnarray}

One needs to be careful when using general expressions like Eq.~(\ref{6pt_amp}) since in some amplitudes an additional term, \textit{i.e.} an $X_i$, exists in the gluon case ($h=1$), compared to the matter amplitudes.

Using our six- and seven-point NMHV amplitudes we have made explicit checks
of Eq.~(\ref{BCJ_matter}), \textit{i.e.} of non-trivial matter relations like
\begin{eqnarray}
m_6(1^-,2^-,\{4_h^-,5_{\bar{h}}^+,6^+\},3^+) =  -\frac{m_6(1^-,2^-,3^+,4_h^-,5_{\bar{h}}^+,6^+)s_{34}(s_{245}+s_{56}+s_{15})s_{16}}{s_{24}s_{245}s_{2456}}&& \nonumber \\
+ (-1)^{2h}\frac{m_6(1^-,2^-,3^+,6^+,5_{\bar{h}}^+,4_h^-)s_{14}(s_{245}+s_{35}+s_{56})(s_{2456}+s_{36})}{s_{24}s_{245}s_{2456}}&& \nonumber \\
+ \frac{m_6(1^-,2^-,3^+,6^+,4_h^-,5_{\bar{h}}^+)(s_{34}+s_{46})s_{15}(s_{2456}+s_{36})}{s_{24}s_{245}s_{2456}}&& \nonumber \\
- (-1)^{2h}\frac{m_6(1^-,2^-,3^+,5_{\bar{h}}^+,4_h^-,6^+)(s_{14}+s_{46})s_{35}s_{16}}{s_{24}s_{245}s_{2456}}&& \nonumber \\
+ \frac{m_6(1^-,2^-,3^+,4_h^-,6^+,5_{\bar{h}}^+)s_{34}s_{15}(s_{2456}+s_{36})}{s_{24}s_{245}s_{2456}}&& \nonumber \\
- (-1)^{2h}\frac{m_6(1^-,2^-,3^+,5_{\bar{h}}^+,6^+,4_h^-)s_{14}s_{35}s_{16}}{s_{24}s_{245}s_{2456}},&&
\end{eqnarray}
and in all our cases we have found Eq.~(\ref{BCJ_matter}) to be satisfied. This verifies our extended relations in a series of highly non-trivial cases.
 
Due to Eq.~(\ref{quark_fermion}) they also hold for the NMHV quark case
$m_n(1_{\bar{q}},2_q,3,\ldots,n)$, with (at least) $n=6$ or 7.

Since the relations are satisfied for MHV amplitudes
independently of the SWI factor (at least when the order of the fermions are kept fixed), one might imagine something similarly to be true for NMHV
amplitudes, \textit{i.e.} valid for a wide range of $h$ values in the $(a_i)^{2-2h}$ factors. However,
in our representation of the amplitudes this was not the case. More
specifically, the relations were satisfied for only very specific values of
$h$, namely $h=1,\nicefrac{1}{2},0,-\nicefrac{1}{2}$ (corresponding to gluons, fermions, scalars
and anti-fermions).

\section{Conclusions}
In this paper we have extended new gluonic relations to
include amplitudes with matter, \textit{i.e.} amplitudes where two
of the particles is a pair of either adjoint fermions,
scalars or quarks. The remaining $n-2$ particles were taken to be
gluons. For MHV amplitudes the extension relied on the simple SWI
structure, multiplying the usual gluon amplitudes by a
factor.

For six- and seven-points we used the BCFW recursion relation to
obtain NMHV matter amplitudes. Using these, explicit checks of the matter relations, Eq.~(\ref{BCJ_matter}), was performed, and in all our cases we found the relations to be satisfied.

The natural extension of the BCJ-relations suggest that the conjectures in \cite{Bern:2008qj} are valid in a very general setting. However, we postpone considerations concerning Eq.~(\ref{BCJ_matter}) for more than two matter particles to future work. 

The importance of understanding the generality of the new relations is not confined to mere tree-level calculations. By mean of unitarity cuts loop amplitudes can be constructed from tree-level amplitudes \cite{Bern:1994zx,Bern:1994cg} (see also \cite{Bern:2007dw}). Considering supersymmetric theories this involve sums over all particle types and physical states that can propagate on the cut lines. It is thus essential for such computations to extend the newly discovered BCJ-relations to these non-gluonic relations.


\section*{Acknowledgements}
We would like to thank N. E. J. Bjerrum-Bohr and P. H. Damgaard for helpful discussions.
	
\appendix
\section{Conventions and notation}
\subsection{Gauge-theory color structure}
Consider an $SU(N_c)$ gauge-theory. When all external particles are in the adjoint representation, the full color-dressed tree amplitude can be written as
\begin{eqnarray}
\mathcal{A}_n^{\mathrm{tree}}\!\!\! & = &\!\!\! g^{n-2}\sum_{\sigma(2,3,\ldots,n)}Tr[T^{a_1}T^{a_{\sigma(2)}}\dots{}T^{a_{\sigma(n)}}] A_n(1,\sigma(2),\ldots,\sigma(n)),
\label{adj_tree}
\end{eqnarray}
where $A_n$ is the partial amplitudes, $T^{a_i}$ the generators of the gauge group, and the sum runs over all permutations of leg $2,3,\ldots,n$ (equivalent to all non-cyclic permutations of leg $1,2,\ldots,n$). We use the shorthand notation $i$ for momentum $p_i$ and have suppressed the helicities.

Although we have used $A_n$ in Eq.~(\ref{adj_tree}) we will in general distinguish between purely gluonic amplitudes, $A_n$, and amplitudes containing matter particles, $m_n$.

The gauge group algebra imply a number of  well known relations between partial amplitudes, see, for instance, \cite{Mangano:1990by}.


Including a quark-antiquark pair, and keeping $n-2$ gluons, the full color-dressed tree amplitude can be decomposed as
\begin{eqnarray}
\mathcal{M}_n^{\mathrm{tree}} \!\!\!& = &\!\!\! g^{n-2}\sum_{\sigma(3,4,\ldots,n)}(T^{a_{\sigma(3)}}T^{a_{\sigma(4)}}\dots{}T^{a_{\sigma(n)}})_{i_2}^{\phantom{i_2}\bar{\jmath}_1}  m_n(1_{\bar{q}},2_q,\sigma(3),\ldots,\sigma(n)),
\label{quark_decom}
\end{eqnarray}
where $i_2$ and $\bar{\jmath}_1$ are the index of the quark and antiquark in the fundamental representation. Note that the quark and antiquark is not part of the permutations in Eq.~(\ref{quark_decom}).
\subsection{Spinor products}
In connection with SWI and MHV amplitudes we use the spinor helicity formalism \cite{Mangano:1990by,Dixon:1996wi}. Actually we only use the spinor products, which will be defined here.

Let $u(p)$ be a massless four-dimensional Dirac spinor, \textit{i.e.}
\begin{eqnarray}
p\cdot\gamma u(p) = 0, \qquad p^2 = 0.
\end{eqnarray}
We always use units with $\hbar = c = 1$ and Minkowski space. Define the two helicity states of $u(p)$ by the two chiral projections
\begin{eqnarray}
u_{\pm}(p) \equiv \frac{1}{2}(1\pm\gamma_5)u(p),
\end{eqnarray}
and introduce the notation
\begin{eqnarray}
|p\pm \rangle \equiv u_{\pm}(p), \qquad \langle p\pm | \equiv \overline{u_{\pm}(p)},
\end{eqnarray}
where $\overline{u_{\pm}(p)} \equiv u_{\pm}(p)\gamma^0$. The basic spinor products are then defined as
\begin{eqnarray}
\langle pq \rangle \equiv \langle p- | q+ \rangle = \overline{u_{-}(p)}u_+(q), \qquad 
 [ pq ] \equiv \langle p+ | q- \rangle = \overline{u_{+}(p)}u_-(q),
\end{eqnarray}
which are related by complex conjugation $\langle pq \rangle^* = [qp]$.

When working with several particles of different momenta $p_i$, we use the shorthand notation
\begin{eqnarray}
\langle p_ip_j \rangle \equiv \langle ij \rangle, \qquad [p_ip_j] \equiv [ij].
\end{eqnarray}

Note that in Eq.~(\ref{6pt_amp}) we use the notation
\begin{eqnarray}
\langle a|b+c|d\rbrack \equiv \langle ab\rangle \lbrack bd\rbrack + \langle ac\rangle \lbrack cd\rbrack.
\end{eqnarray}

\section{BCJ-relations for six- and seven-point amplitudes}
We give the explicit expressions for the six- and seven-point relations generated from Eq.~(\ref{BCJ}). The matter relations can be obtained by replacing the gluon amplitudes, $A_n$, with the matter amplitudes, $m_n$, and multiplying each of them with the sign-function from Eq.~(\ref{BCJ_matter}).
 
The three classes of BCJ-relations for gluonic six-point amplitudes are \cite{Bern:2008qj}
\begin{eqnarray}
A_6(1,2,\{4\},3,\{5,6\}) \!\!\!&=&\!\!\! \frac{A_6(1,2,3,4,5,6)(s_{14}+s_{46}+s_{45})}{s_{24}} \nonumber \\
&&\!\!\!+ \frac{A_6(1,2,3,5,4,6)(s_{14}+s_{46})}{s_{24}} \nonumber \\
&&\!\!\! + \frac{A_6(1,2,3,5,6,4)s_{14}}{s_{24}},
\label{first_six_point} \\ \nonumber \\ \nonumber \\
A_6(1,2,\{4,5\},3,\{6\}) \!\!\!& = &\!\!\! -\frac{A_6(1,2,3,4,5,6)s_{34}(s_{15}+s_{56})}{s_{24}s_{245}}\nonumber \\
&&\!\!\!-\frac{A_6(1,2,3,4,6,5)s_{34}s_{15}}{s_{24}s_{245}} \nonumber \\
&&\!\!\!- \frac{A_6(1,2,3,6,4,5)(s_{34}+s_{46})s_{15}}{s_{24}s_{245}} \nonumber \\
&&\!\!\! - \frac{A_6(1,2,3,5,4,6)(s_{14}+s_{46})(s_{245}+s_{35})}{s_{24}s_{245}} \nonumber \\
&&\!\!\! - \frac{A_6(1,2,3,5,6,4)s_{14}(s_{245}+s_{35})}{s_{24}s_{245}} \nonumber \\
&&\!\!\!- \frac{A_6(1,2,3,6,5,4)s_{14}(s_{245}+s_{35}+s_{56})}{s_{24}s_{245}},
\label{second_six_point}
\end{eqnarray}
\begin{eqnarray}
\phantom{AAAA,}\!A_6(1,2,\{4,5,6\},3) \!\!\!& = &\!\!\! -\frac{A_6(1,2,3,4,5,6)s_{34}(s_{245}+s_{56}+s_{15})s_{16}}{s_{24}s_{245}s_{2456}} \nonumber \\
&&\!\!\!+ \frac{A_6(1,2,3,4,6,5)s_{34}s_{15}(s_{2456}+s_{36})}{s_{24}s_{245}s_{2456}} \nonumber \\
&&\!\!\!+ \frac{A_6(1,2,3,6,4,5)(s_{34}+s_{46})s_{15}(s_{2456}+s_{36})}{s_{24}s_{245}s_{2456}} \nonumber \\
&&\!\!\!- \frac{A_6(1,2,3,5,4,6)(s_{14}+s_{46})s_{35}s_{16}}{s_{24}s_{245}s_{2456}} \nonumber \\
&&\!\!\!- \frac{A_6(1,2,3,5,6,4)s_{14}s_{35}s_{16}}{s_{24}s_{245}s_{2456}}\nonumber \\
&&\!\!\!+ \frac{A_6(1,2,3,6,5,4)s_{14}(s_{245}+s_{35}+s_{56})(s_{2456}+s_{36})}{s_{24}s_{245}s_{2456}}.
\label{third_six_point}
\end{eqnarray}
\\ 
We get the following four classes of BCJ-relations for gluonic seven-point amplitudes
\begin{eqnarray}
A_7(1,2,\{4\},3,\{5,6,7\}) \!\!\!& = &\!\!\! \frac{A_7(1,2,3,4,5,6,7)(s_{45}+s_{46}+s_{47}+s_{41})}{s_{24}} \nonumber \\
&&\!\!\!+ \frac{A_7(1,2,3,5,4,6,7)(s_{46}+s_{47}+s_{41})}{s_{24}} \nonumber \\
&&\!\!\! +\frac{A_7(1,2,3,5,6,4,7)(s_{47}+s_{41})}{s_{24}} \nonumber \\
&& \!\!\!+ \frac{A_7(1,2,3,5,6,7,4)s_{41}}{s_{24}},
\label{first_seven} \\ \nonumber \\ \nonumber \\
A_7(1,2,\{4,5\},3,\{6,7\}) \!\!\!& = &\!\!\! -\frac{A_7(1,2,3,4,5,6,7)s_{43}(s_{56}+s_{57}+s_{51})}{s_{24}s_{245}} \nonumber \\
&&\!\!\!- \frac{A_7(1,2,3,5,4,6,7)(s_{46}+s_{47}+s_{41})(s_{53}+s_{245})}{s_{24}s_{245}} \nonumber \\
&&\!\!\! -\frac{A_7(1,2,3,4,6,5,7)s_{43}(s_{57}+s_{51})}{s_{24}s_{245}} \nonumber \\
&&\!\!\! -\frac{A_7(1,2,3,4,6,7,5)s_{43}s_{51}}{s_{24}s_{245}} \nonumber \\
&&\!\!\!-\frac{A_7(1,2,3,6,4,7,5)(s_{43}+s_{46})s_{51}}{s_{24}s_{245}} \nonumber \\
&&\!\!\! - \frac{A_7(1,2,3,6,7,4,5)(s_{43}+s_{46}+s_{47})s_{51}}{s_{24}s_{245}} \nonumber \\
&&\!\!\! -\frac{A_7(1,2,3,6,7,5,4)s_{41}(s_{53}+s_{56}+s_{57}+s_{245})}{s_{24}s_{245}} \nonumber \\
&&\!\!\!-\frac{A_7(1,2,3,6,5,7,4)s_{41}(s_{53}+s_{56}+s_{245})}{s_{24}s_{245}} \nonumber \\
&&\!\!\!-\frac{A_7(1,2,3,5,6,7,4)s_{41}(s_{53}+s_{245})}{s_{24}s_{245}} \nonumber \\
&&\!\!\! -\frac{A_7(1,2,3,5,6,4,7)(s_{47}+s_{41})(s_{53}+s_{245})}{s_{24}s_{245}} \nonumber \\
&&\!\!\! -\frac{A_7(1,2,3,6,4,5,7)(s_{43}+s_{46})(s_{57}+s_{51})}{s_{24}s_{245}} \nonumber \\
&&\!\!\! - \frac{A_7(1,2,3,6,5,4,7)(s_{47}+s_{41})(s_{53}+s_{56}+s_{245})}{s_{24}s_{245}},
\end{eqnarray}
\begin{eqnarray}
A_7(1,2,\{4,5,6\},3,\{7\}) \!\!\!&=&\!\!\! -\frac{A_7(1,2,3,4,5,6,7)s_{43}(s_{56}+s_{57}+s_{51}+s_{245})(s_{67}+s_{61})}{s_{24}s_{245}s_{2456}} \nonumber \\
&&\!\!\!+\frac{A_7(1,2,3,6,7,5,4)s_{41}(s_{53}+s_{56}+s_{57}+s_{245})(s_{63}+s_{2456})}{s_{24}s_{245}s_{2456}} \nonumber \\
&&\!\!\!-\frac{A_7(1,2,3,7,4,5,6)(s_{43}+s_{47})(s_{56}+s_{51}+s_{245})s_{61}}{s_{24}s_{245}s_{2456}} \nonumber \\
&&\!\!\!+\frac{A_7(1,2,3,6,5,4,7)(s_{47}+s_{41})(s_{53}+s_{56}+s_{245})(s_{63}+s_{2456})}{s_{24}s_{245}s_{2456}} \nonumber \\
&&\!\!\!-\frac{A_7(1,2,3,5,6,4,7)(s_{47}+s_{41})s_{53}(s_{67}+s_{61})}{s_{24}s_{245}s_{2456}} \nonumber \\
&&\!\!\!+\frac{A_7(1,2,3,4,6,5,7)s_{43}(s_{57}+s_{51})(s_{63}+s_{2456})}{s_{24}s_{245}s_{2456}}\nonumber \\
&&\!\!\!+\frac{A_7(1,2,3,6,4,5,7)(s_{43}+s_{46})(s_{57}+s_{51})(s_{63}+s_{2456})}{s_{24}s_{245}s_{2456}} \nonumber \\
&&\!\!\!-\frac{A_7(1,2,3,5,4,6,7)(s_{46}+s_{47}+s_{41})s_{53}(s_{67}+s_{61})}{s_{24}s_{245}s_{2456}}
\nonumber \\
&&\!\!\!-\frac{A_7(1,2,3,4,5,7,6)s_{43}(s_{57}+s_{56}+s_{51}+s_{245})s_{61}}{s_{24}s_{245}s_{2456}} \nonumber \\
&&\!\!\!+\frac{A_7(1,2,3,6,5,7,4)s_{41}(s_{53}+s_{56}+s_{245})(s_{63}+s_{2456})}{s_{24}s_{245}s_{2456}}\nonumber \\
&&\!\!\!-\frac{A_7(1,2,3,4,7,5,6)s_{43}(s_{56}+s_{51}+s_{245})s_{61}}{s_{24}s_{245}s_{2456}} \nonumber \\
&&\!\!\!-\frac{A_7(1,2,3,7,5,4,6)(s_{46}+s_{41})(s_{53}+s_{57})s_{61}}{s_{24}s_{245}s_{2456}}\nonumber \\
&&\!\!\!+\frac{A_7(1,2,3,7,4,6,5)(s_{43}+s_{47})s_{51}(s_{63}+s_{67}+s_{2456})}{s_{24}s_{245}s_{2456}} \nonumber \\
&&\!\!\!-\frac{A_7(1,2,3,7,5,6,4)s_{41}(s_{53}+s_{57})s_{61}}{s_{24}s_{245}s_{2456}}\nonumber \\
&&\!\!\!+\frac{A_7(1,2,3,7,6,4,5)(s_{43}+s_{47}+s_{46})s_{51}(s_{63}+s_{67}+s_{2456})}{s_{24}s_{245}s_{2456}} \nonumber \\
&&\!\!\!-\frac{A_7(1,2,3,5,6,7,4)s_{41}s_{53}(s_{67}+s_{61})}{s_{24}s_{245}s_{2456}}\nonumber \\
&&\!\!\!+\frac{A_7(1,2,3,6,4,7,5)(s_{43}+s_{46})s_{51}(s_{63}+s_{2456})}{s_{24}s_{245}s_{2456}} \nonumber \\
&&\!\!\!-\frac{A_7(1,2,3,5,4,7,6)(s_{47}+s_{46}+s_{41})s_{53}s_{61}}{s_{24}s_{245}s_{2456}}\nonumber \\
&&\!\!\!+\frac{A_7(1,2,3,4,6,7,5)s_{43}s_{51}(s_{63}+s_{2456})}{s_{24}s_{245}s_{2456}} \nonumber \\
&&\!\!\!-\frac{A_7(1,2,3,5,7,4,6)(s_{46}+s_{41})s_{53}s_{61}}{s_{24}s_{245}s_{2456}}\nonumber \\
&&\!\!\!+\frac{A_7(1,2,3,4,7,6,5)s_{43}s_{51}(s_{63}+s_{67}+s_{2456})}{s_{24}s_{245}s_{2456}} \nonumber \\
&&\!\!\!-\frac{A_7(1,2,3,5,7,6,4)s_{41}s_{53}s_{61}}{s_{24}s_{245}s_{2456}} \nonumber \\
&&\!\!\!+\frac{A_7(1,2,3,6,7,4,5)(s_{43}+s_{46}+s_{47})s_{51}(s_{63}+s_{2456})}{s_{24}s_{245}s_{2456}} \nonumber \\
&&\!\!\!+\frac{A_7(1,2,3,7,6,5,4)s_{41}(s_{53}+s_{57}+s_{56}+s_{245})(s_{63}+s_{67}+s_{2456})}{s_{24}s_{245}s_{2456}}, \nonumber \\
\end{eqnarray}
\begin{eqnarray}
A_7(1,2,\{4,5,6,7\},3) \!\!\!& =&\!\!\! -\frac{A_7(1,2,3,4,5,6,7)s_{43}(s_{56}+s_{57}+s_{51}+s_{245})(s_{67}+s_{61}+s_{2456})s_{71}}{s_{24}s_{245}s_{2456}s_{24567}} \nonumber \\
&&\!\!\!-\frac{A_7(1,2,3,5,4,6,7)(s_{46}+s_{47}+s_{41})s_{53}(s_{67}+s_{61}+s_{2456})s_{71}}{s_{24}s_{245}s_{2456}s_{24567}}\nonumber \\
&&\!\!\!+\frac{A_7(1,2,3,7,5,6,4)s_{41}(s_{53}+s_{57})s_{61}(s_{73}+s_{24567})}{s_{24}s_{245}s_{2456}s_{24567}} \nonumber \\
&&\!\!\!+\frac{A_7(1,2,3,7,5,4,6)(s_{46}+s_{41})(s_{53}+s_{57})s_{61}(s_{73}+s_{24567})}{s_{24}s_{245}s_{2456}s_{24567}}\nonumber \\ &&\!\!\!+\frac{A_7(1,2,3,6,7,5,4)s_{41}(s_{53}+s_{56}+s_{57}+s_{245})s_{63}s_{71}}{s_{24}s_{245}s_{2456}s_{24567}} \nonumber \\
&&\!\!\!+\frac{A_7(1,2,3,6,5,4,7)(s_{47}+s_{41})(s_{53}+s_{56}+s_{245})s_{63}s_{71}}{s_{24}s_{245}s_{2456}s_{24567}}\nonumber \\
&&\!\!\!-\frac{A_7(1,2,3,5,6,4,7)(s_{47}+s_{41})s_{53}(s_{67}+s_{61}+s_{2456})s_{71}}{s_{24}s_{245}s_{2456}s_{24567}} \nonumber \\
&&\!\!\!+\frac{A_7(1,2,3,5,7,4,6)(s_{46}+s_{41})s_{53}s_{61}(s_{73}+s_{24567})}{s_{24}s_{245}s_{2456}s_{24567}} \nonumber \\
&&\!\!\!-\frac{A_7(1,2,3,4,7,6,5)s_{43}s_{51}(s_{63}+s_{67}+s_{2456})(s_{73}+s_{24567})}{s_{24}s_{245}s_{2456}s_{24567}} \nonumber \\
&&\!\!\!+\frac{A_7(1,2,3,6,5,7,4)s_{41}(s_{53}+s_{56}+s_{245})s_{63}s_{71}}{s_{24}s_{245}s_{2456}s_{24567}}\nonumber \\
&&\!\!\!+\frac{A_7(1,2,3,4,7,5,6)s_{43}(s_{56}+s_{51}+s_{245})s_{61}(s_{73}+s_{24567})}{s_{24}s_{245}s_{2456}s_{24567}} \nonumber \\
&&\!\!\!+\frac{A_7(1,2,3,5,4,7,6)(s_{47}+s_{46}+s_{41})s_{53}s_{61}(s_{73}+s_{24567})}{s_{24}s_{245}s_{2456}s_{24567}}\nonumber \\
&&\!\!\!+\frac{A_7(1,2,3,4,6,7,5)s_{43}s_{51}s_{63}s_{71}}{s_{24}s_{245}s_{2456}s_{24567}} \nonumber \\
&&\!\!\!+\frac{A_7(1,2,3,5,7,6,4)s_{41}s_{53}s_{61}(s_{73}+s_{24567})}{s_{24}s_{245}s_{2456}s_{24567}}\nonumber \\
&&\!\!\!+\frac{A_7(1,2,3,6,7,4,5)(s_{43}+s_{46}+s_{47})s_{51}s_{63}s_{71}}{s_{24}s_{245}s_{2456}s_{24567}} \nonumber \\
&&\!\!\!+\frac{A_7(1,2,3,4,6,5,7)s_{43}(s_{57}+s_{51})s_{63}s_{71}}{s_{24}s_{245}s_{2456}s_{24567}}\nonumber \\
&&\!\!\!+\frac{A_7(1,2,3,6,4,5,7)(s_{43}+s_{46})(s_{57}+s_{51})s_{63}s_{71}}{s_{24}s_{245}s_{2456}s_{24567}} \nonumber \\
&&\!\!\!-\frac{A_7(1,2,3,5,6,7,4)s_{41}s_{53}(s_{67}+s_{61}+s_{2456})s_{71}}{s_{24}s_{245}s_{2456}s_{24567}}\nonumber \\
&&\!\!\!+\frac{A_7(1,2,3,6,4,7,5)(s_{43}+s_{46})s_{51}s_{63}s_{71}}{s_{24}s_{245}s_{2456}s_{24567}} \nonumber \\
&&\!\!\!-\frac{A_7(1,2,3,7,6,5,4)s_{41}(s_{53}+s_{57}+s_{56}+s_{245})(s_{63}+s_{67}+s_{2456})(s_{73}+s_{24567})}{s_{24}s_{245}s_{2456}s_{24567}} \nonumber \\ 
&&\!\!\!-\frac{A_7(1,2,3,7,6,4,5)(s_{43}+s_{47}+s_{46})s_{51}(s_{63}+s_{67}+s_{2456})(s_{73}+s_{24567})}{s_{24}s_{245}s_{2456}s_{24567}} \nonumber \\
&&\!\!\!+\frac{A_7(1,2,3,7,4,5,6)(s_{43}+s_{47})(s_{56}+s_{51}+s_{245})s_{61}(s_{73}+s_{24567})}{s_{24}s_{245}s_{2456}s_{24567}} \nonumber \\
&&\!\!\!-\frac{A_7(1,2,3,7,4,6,5)(s_{43}+s_{47})s_{51}(s_{63}+s_{67}+s_{2456})(s_{73}+s_{24567})}{s_{24}s_{245}s_{2456}s_{24567}} \nonumber \\
&&\!\!\!+\frac{A_7(1,2,3,4,5,7,6)s_{43}(s_{57}+s_{56}+s_{51}+s_{245})s_{61}(s_{73}+s_{24567})}{s_{24}s_{245}s_{2456}s_{24567}}. 
\end{eqnarray}

\bibliographystyle{unsrt}
\bibliography{article}

\begin{thebibliography}{10}

\bibitem{Witten:2003nn}
Edward Witten.
\newblock {Perturbative gauge theory as a string theory in twistor space}.
\newblock {\em Commun. Math. Phys.}, 252:189--258, 2004.

\bibitem{Cachazo:2004kj}
Freddy Cachazo, Peter Svrcek, and Edward Witten.
\newblock {MHV vertices and tree amplitudes in gauge theory}.
\newblock {\em JHEP}, 09:006, 2004.

\bibitem{Parke:1986gb}
Stephen~J. Parke and T.~R. Taylor.
\newblock {An Amplitude for $n$ Gluon Scattering}.
\newblock {\em Phys. Rev. Lett.}, 56:2459, 1986.

\bibitem{Berends:1987me}
Frits~A. Berends and W.~T. Giele.
\newblock {Recursive Calculations for Processes with n Gluons}.
\newblock {\em Nucl. Phys.}, B306:759, 1988.

\bibitem{Britto:2004ap}
Ruth Britto, Freddy Cachazo, and Bo~Feng.
\newblock {New Recursion Relations for Tree Amplitudes of Gluons}.
\newblock {\em Nucl. Phys.}, B715:499--522, 2005.

\bibitem{Britto:2005fq}
Ruth Britto, Freddy Cachazo, Bo~Feng, and Edward Witten.
\newblock {Direct Proof Of Tree-Level Recursion Relation In Yang- Mills
  Theory}.
\newblock {\em Phys. Rev. Lett.}, 94:181602, 2005.

\bibitem{Benincasa:2007xk}
Paolo Benincasa and Freddy Cachazo.
\newblock {Consistency Conditions on the S-Matrix of Massless Particles}.
\newblock 2007.

\bibitem{Schuster:2008nh}
Philip~C. Schuster and Natalia Toro.
\newblock {Constructing the Tree-Level Yang-Mills S-Matrix Using Complex
  Factorization}.
\newblock 2008.

\bibitem{Mangano:1990by}
Michelangelo~L. Mangano and Stephen~J. Parke.
\newblock {Multi-Parton Amplitudes in Gauge Theories}.
\newblock {\em Phys. Rept.}, 200:301--367, 1991.

\bibitem{Dixon:1996wi}
Lance~J. Dixon.
\newblock {Calculating scattering amplitudes efficiently}.
\newblock 1996.

\bibitem{Kleiss:1988ne}
Ronald Kleiss and Hans Kuijf.
\newblock {Multi - Gluon Cross-Sections And Five Jet Production At Hadron
  Colliders}.
\newblock {\em Nucl. Phys.}, B312:616, 1989.

\bibitem{DelDuca:1999rs}
Vittorio Del~Duca, Lance~J. Dixon, and Fabio Maltoni.
\newblock {New color decompositions for gauge amplitudes at tree and loop
  level}.
\newblock {\em Nucl. Phys.}, B571:51--70, 2000.

\bibitem{Grisaru:1976vm}
Marcus~T. Grisaru, H.~N. Pendleton, and P.~van Nieuwenhuizen.
\newblock {Supergravity and the S Matrix}.
\newblock {\em Phys. Rev.}, D15:996, 1977.

\bibitem{Grisaru:1977px}
Marcus~T. Grisaru and H.~N. Pendleton.
\newblock {Some Properties of Scattering Amplitudes in Supersymmetric
  Theories}.
\newblock {\em Nucl. Phys.}, B124:81, 1977.

\bibitem{Parke:1985pn}
Stephen~J. Parke and T.~R. Taylor.
\newblock {Perturbative QCD Utilizing Extended Supersymmetry}.
\newblock {\em Phys. Lett.}, B157:81, 1985.

\bibitem{Bern:2008qj}
Z.~Bern, J.~J.~M. Carrasco, and Henrik Johansson.
\newblock {New Relations for Gauge-Theory Amplitudes}.
\newblock {\em Phys. Rev.}, D78:085011, 2008.

\bibitem{Nair:1988bq}
V.~P. Nair.
\newblock {A Current Algebra For Some Gauge Theory Amplitudes}.
\newblock {\em Phys. Lett.}, B214:215, 1988.

\bibitem{Mangano:1987kp}
Michelangelo~L. Mangano and Stephen~J. Parke.
\newblock {Quark - Gluon Amplitudes in the Dual Expansion}.
\newblock {\em Nucl. Phys.}, B299:673, 1988.

\bibitem{Bidder:2005in}
Steven~J. Bidder, David~C. Dunbar, and Warren~B. Perkins.
\newblock {Supersymmetric Ward identities and NMHV amplitudes involving
  gluinos}.
\newblock {\em JHEP}, 08:055, 2005.

\bibitem{Britto:2005ha}
Ruth Britto, Evgeny Buchbinder, Freddy Cachazo, and Bo~Feng.
\newblock {One-loop amplitudes of gluons in SQCD}.
\newblock {\em Phys. Rev.}, D72:065012, 2005.

\bibitem{Georgiou:2004wu}
George Georgiou and Valentin~V. Khoze.
\newblock {Tree amplitudes in gauge theory as scalar MHV diagrams}.
\newblock {\em JHEP}, 05:070, 2004.

\bibitem{Glover:2008tu}
E.~W.~Nigel Glover, Valentin~V. Khoze, and Ciaran Williams.
\newblock {Component MHV amplitudes in N=2 SQCD and in N=4 SYM at one loop}.
\newblock {\em JHEP}, 08:033, 2008.

\bibitem{Zhu:1980sz}
Dong-pei Zhu.
\newblock {Zeros In Scattering Amplitudes And The Structure Of Nonabelian Gauge
  Theories}.
\newblock {\em Phys. Rev.}, D22:2266, 1980.

\bibitem{de_Florian:2006ek}
Daniel de~Florian and Jose Zurita.
\newblock {Seven parton amplitudes from recursion relations}.
\newblock {\em JHEP}, 05:073, 2006.

\bibitem{Bern:1994zx}
Zvi Bern, Lance~J. Dixon, David~C. Dunbar, and David~A. Kosower.
\newblock {One-Loop n-Point Gauge Theory Amplitudes, Unitarity and Collinear
  Limits}.
\newblock {\em Nucl. Phys.}, B425:217--260, 1994.

\bibitem{Bern:1994cg}
Zvi Bern, Lance~J. Dixon, David~C. Dunbar, and David~A. Kosower.
\newblock {Fusing gauge theory tree amplitudes into loop amplitudes}.
\newblock {\em Nucl. Phys.}, B435:59--101, 1995.

\bibitem{Bern:2007dw}
Zvi Bern, Lance~J. Dixon, and David~A. Kosower.
\newblock {On-Shell Methods in Perturbative QCD}.
\newblock {\em Annals Phys.}, 322:1587--1634, 2007.

\end{thebibliography}

\end{document}